\documentclass[12pt,a4paper]{article}

\usepackage{amsmath}
\def\RR{\mbox{I\hspace{-.15em}R}}
\def\DD{\mbox{I\hspace{-.15em}D}}
\newcommand{\ds}{\displaystyle}
\input{rotate}
\input epsf

\begin{document}


\title{Scaling transformation and probability distributions for financial time series}

\author{Marc-Etienne BRACHET \footnote{Laboratoire de Physique Statistique, CNRS URA 1306, ENS, 24 Rue Lhomond, 75231 Paris Cedex 05, France} \and Erik TAFLIN 
\footnote{Direction Scientifique, AXA-UAP, 9 Place Vend\^{o}me, 
75052 Paris Cedex 01, France} \and Jean Marcel TCHEOU$^{\ast \dagger}$}

\date{December 1997}
\maketitle

\begin{abstract}
The price of financial assets are, since \cite{Bachelier}, considered to 
be described by a (discrete or continuous) time sequence of random variables,
i.e a stochastic process.
Sharp scaling exponents or unifractal behavior of such processes has been 
reported in several works \cite{Mandel1} \cite{Peters} \cite{Stan} 
\cite{Evertz} \cite{Bouchaud}.
In this letter we investigate the question of scaling transformation of price
processes by establishing a new connexion between non-linear group theoretical
methods and multifractal methods developed in mathematical physics. Using two sets of financial chronological time series, we show  
that the scaling transformation is a non-linear group action on the moments of the price increments. Its 
linear part has a spectral decomposition that puts in evidence a
multifractal behavior of the price increments.
\end{abstract}

\section{Introduction}
One of the pillars of modern physics is the covariance of theories under 
certain group actions. What is particular to a given application,
such as initial and boundary conditions usually breaks the symmetries of 
the theory. The symmetry group of observed data is therefore usually much smaller than the covariance group of the theory. An example is hydrodynamics where the equations are 
invariant under space-time translation and scaling, but where 
the solutions are not, in general, invariant. For theories that are covariant 
under scaling (to be specific, we can think of Navier-Stokes or Korteweg-De Vries equation) the situation is clear: the scaling properties, such as the spectral decomposition of the 
solution at each time are given by the scaling properties of the initial 
condition, the boundary conditions and external forces. 
The study of scaling transformation properties in the domain of economics and
finance is more complicated because the evolution equations
(or even the theory) governing the dynamics are largely unknown. It is thus not possible, in this case,
to separate these scaling properties into a general property of an
underlying theory and into what is particular to the situation under
study. 
The observed financial chronological data results at least to some extent, of the particularities
of each market and not only from a general abstract dynamics. Therefore there is
no {\it a priori} reason to expect the data to exhibit simple properties under 
scaling transformation. Keeping this simple observation in mind, we will base 
our analysis of the scaling
transformation on methods adapted to physical systems with complex behavior:
\smallskip \\
i) Multifractal analysis of fully turbulent systems introduced in \cite{FP}
and within that approach further developed inversion techniques 
developed in \cite{Br.Tch1} (see also \cite{Br.Tch}). \\
ii) Non-linear group representation theory developed in \cite{FPS} and applied
to many non-linear evolution equations in mathematical physics (see \cite{FST} for
recent contributions).
\smallskip \\
We apply these methods to two sets of financial chronological series :

1) Foreign exchange rate DM/$\$$: The data set provided by Olsen and Associates
 contains worldwide 1,472,241 bid-ask quotes for US dollar-German mark exchanges rates from 1 October 1992 until 30 September 1993.
Tick by tick data are irregularly spaced in recorded time. 
To obtain price values at
 a regular time, we use linear interpolation between the two recorded time
that immediately precede and follow the regular time. We obtain in this way, for a regular time of $15$ seconds, $1,059,648$ data.
Our study focuses on the average price which is the mean of the bid and ask price.

2) Stock index CAC 40: The data set provided by the ``Soci\'et\'e de Bourse Fran\c{c}aise'' contains $1,045,890$ quotes of the CAC 40 index from 3 January 1993 until 31 December 1996.
Tick by tick data are regularly recorded every $30$ seconds,
during opening hours 
(everyday from 10 am until 5 pm except week ends and national holidays). 
Our data base consist of the daily registers to which a constant has been subtracted such that 
the value at 10am is equal to the value of the previous day at 5pm.
The subtracted jump process (with jumps at fixed times) can be analyzed on its own.
This separation allows for a finer analysis of the rest of the process.

Using these data sets, we obtained three new results. First, the scaling
transformation of the moments of the
observed probability distribution is a non-linear representation that is well approximated  
by a linear representation, for small scaling parameters. This linear representation turns out to be diagonal.
Secondly, the function of the order of the moment, defined by the spectrum of the generator
is (non trivially) concave. This shows, by definition \cite{FP}, that the data are multifractal. 
Note that the concavity in the case of FX market (DM/$\$$) can partially be
deduced from \cite{ghash}
and is confirmed, independently of our work, by \cite{Mandel2}.
Our third new result is an explicit expression of the family of probability distributions of price increments corresponding to different time increments.

For larger values of the scaling parameter, the linear approximation breaks down
and the non linear terms of the representation has to be considered. 
The analysis of this letter can also be applied to the SP 500 index, where the 
results should be compared to the (unifractal) scaling behavior found in \cite{Stan}.
It should also be compared with the, from the point of view of finance, more fundamental approach of
stochastic time transformation (subordinate processes) that were applied to SP500 \cite{Geman1} \cite{Geman2}.
These points are left for future investigations.

\section{The mathematical framework}
We suppose that the financial variable is described by a stochastic process $(u(t))_{t \geq 0}$ such that the set of increments $u(t+\tau)-u(t)$, $\tau \geq 0$ has a well defined transformation property under scaling of the time increment $\tau$, $\tau \mapsto a \: \tau$, $a > 0$.
To avoid complications, irrelevant for the quite crude application reported in this
letter, we suppose that $(u(t))_{t \geq 0}$ is stationary.
Moreover, we will only consider the absolute value $|u(t+\tau)-u(t)|$ of 
increments. Let $w(\tau)=|u(\tau)-u(0)|$, $\tau \geq 0$. This means that for 
each (scaling) $a > 0$, there is a map $T_a$ of the set 
$\mathcal {W}=\{ w(\tau) | \tau \geq 0 \}$ such that $T_a(w(\tau)) = w(a \tau)$.
A group action $T$ of the scaling (dilatation) group $\DD$ (the set of strictly
 positive real numbers) on the set $\mathcal {W}$ is then defined, i.e 
$T_{ab}(x) = T_a(T_b(x))$ and  
$T_e(x)=x$ for $a, b \in \DD, x \in \mathcal {W}$, where $e=1$ is the identity element
 in $\DD$. In the cases under consideration in this letter, it follows from
the observed time series that the estimated probability distribution $p_{w(\tau)}$ in $\RR$ of $w(\tau)$ is different for different $\tau > 0$. This is enough to ensure
the existence of the action $T$, and moreover shows that $T$ gives a group action $\bar{T}$, on the set $\mathcal {M} = \{ p_{w(\tau)} | \tau \geq 0 \}$ of probability distributions, defined by $\bar{T}_a(p_{w(\tau)}) = p_{w(a \tau)}$.

The group action $\bar{T}$ is not linear, in spite of its
appearance. To explicit properties of the scaling action $\bar{T}$, we change the coordinates of the elements in $\mathcal {M}$. As in the case
of fully developed turbulence, we use the moments as coordinates.
For $q \in \mathcal {M}$, let the moment vector be the sequence 
$S(q) = (S_{r}(q))_{r \geq 0}$, where 
$S_r(q) = \int_{0}^{\infty} x^r q(x) \: dx$ and $r \in \RR^{+}$.
Here we suppose that the set $\mathcal {M}$ of probability measures is such that
$S_r(q)$ exists for all orders $r$ and moreover that $q$ is determined by its moments of order $r \in \RR^{+}$ (which is the case if for example the Fourier Transform of elements
in $\mathcal {M}$ are quasi analytic). Let $\mathcal {S}$ be the image (in the space $C(\RR^{+})$ of continuous real functions on $\RR^{+}$) of $\mathcal {M}$ under the coordinate 
transformation $S$. The image $U$ of the group action $\bar{T}$ is 
given by $U_a = S \circ \bar{T}_a \circ S^{-1}$, i.e 
$U_a(S(p_{w(\tau)}))=S(p_{w(a \tau)})$.
In the case $U$ is a linear diagonal representation, it has the form $U_a=U_a^{(1)}$, where for given real numbers $\zeta_r$ with $r \in \RR^{+}$:
\begin{eqnarray}
U_a^{(1)}(m) = (a^{\zeta_r} \: m_r)_{r \geq 0}
\label{eq1}
\end{eqnarray}
for $a \in \DD$ and $m \in C(\RR^{+})$, $m_r$ corresponding to a moment of order $r$.
We note that $\{ \zeta_r | r \in \RR^{+} \}$ is the spectrum of the generator of the representation $U^{(1)}$.
When $U$ is a non linear perturbation of $U^{(1)}$, there are algorithms permitting its construction. However they are outside the scope of this letter \cite{FPS} \cite{FST}. For commodity we denote 
$s_r(\tau) = S_r(p_{w(\tau)})$ which is the r-th component of $U_{\tau}(S(p_{w(1)}))$.
An accurate and explicit approximation of the inverse transformation $S^{-1}$, of the moment vectors $s_r(\tau)$ to probability distribution $p_{w(\tau)}$ has
been developed in \cite{Br.Tch1} \cite{Br.Tch} \cite{theseJMT}. This permits to obtain directly from 
experimental data, an explicit formula for the family 
${\cal M} = \{ p_{w(\tau)} \}_{\tau>0}$
of probabilities.
In fact, for each $\tau \in \RR^{+}$ we can determine an element $p_{w(\tau)}$
by the formulas: 

\begin{eqnarray}
x \: p_{w(\tau)}(x) & = & \bar{p}(\ln x)
\nonumber \\
\alpha(r,\tau) & = & \ds{\frac{d \ln s_r(\tau)}{dr}}
\label{uap7} \\
 \ln \bar{p}(\alpha(r,\tau)) & = & \ln s_r(\tau)-r \ds{\frac{d \ln s_r(\tau)}{dr}} - \ds{\frac{1}
{2}} \ln (2 \pi) - \ds{\frac{1}{2}} \ds{\frac{d^2 \ln s_r(\tau)}{dr^2}} .
\label{uap8}
\end{eqnarray}
where $r \in \RR^{+}$.

\section{Results}
\begin{figure}[tbhp]
\centerline{\epsfxsize=12truecm\epsfbox{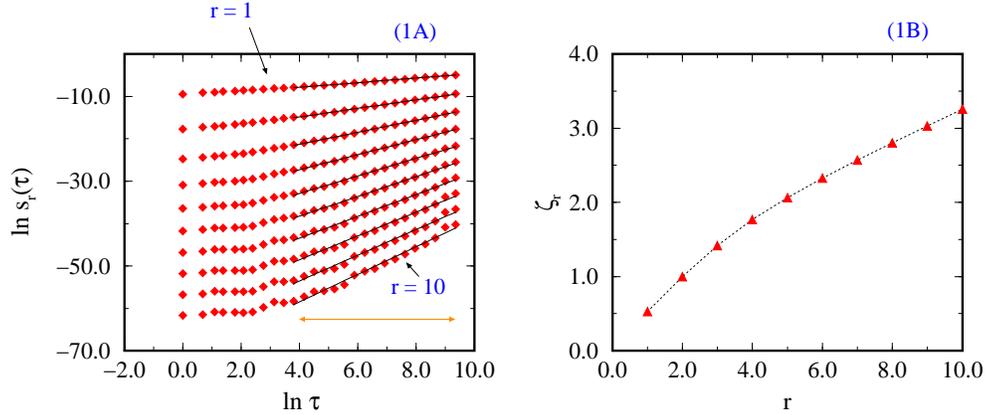}}
\caption{(1A): $\ln s_r(\tau)$ vs $\ln \tau$ for $r=1 \ldots 10$ in the case
of FX DM/$\$$ index.
The scaling law $\ln s_r(\tau) = A_r + \zeta_r \ln \tau$ displayed as straight
 lines for $11 \leq \tau \leq 2896$ minutes (delimited by the orange arrow) gives $\zeta_r$ as the slope.
(1B): $\zeta_r$ determined by scaling law in the case of FX DM/$\$$ index.}
\label{fig1}
\end{figure}

When the representation $U$ is linear it follows from expression 
(\ref{eq1}) that $\ln s_r(\tau) = A_r + \zeta_r \ln \tau$, where $A_r$ and 
$\zeta_r$ are independent of $\tau$. Figure (\ref{fig1}A) 
shows that, in the case of FX DM/$\$$, this is
satisfied, to a good approximation with time increments $\tau$ and moments of 
order $r$ in the interval $11 \leq \tau \leq 2896$ minutes and $1 \leq r \leq 10$.
In contrast, for CAC40 the domain of validity of
\begin{figure}[tbhp]
\centerline{\epsfxsize=12truecm\epsfbox{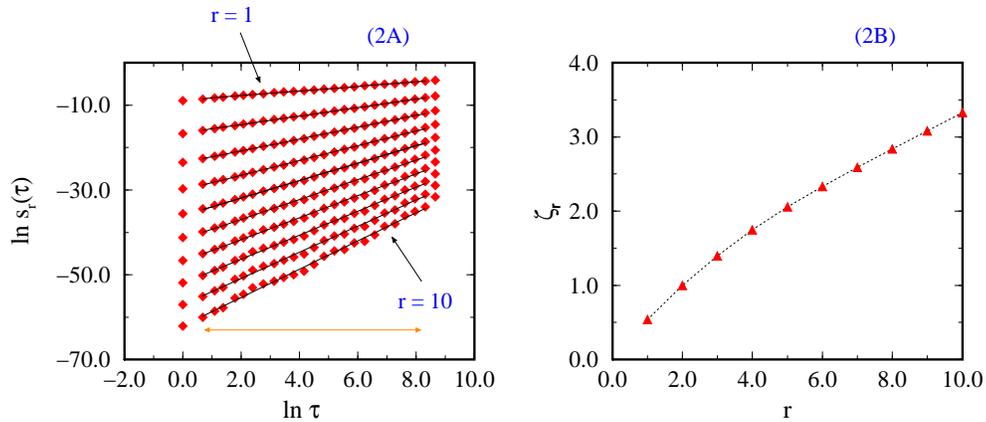}}
\caption{(2A): $\ln s_r(\tau)$ vs $\ln \tau$ for $r=1 \ldots 10$ in the case
of CAC40 index.
The scaling law $\ln s_r(\tau) = A_r + \zeta_r \ln \tau$ displayed as straight
 lines for $1 \leq \tau \leq 2048$ minutes (delimited by the orange arrow) gives
 $\zeta_r$ as the slope.
(2B): $\zeta_r$ determined by scaling law in the case of CAC40 index.}
\label{fig2}
\end{figure}
the linear approximation also contains
the small values of $\tau$:  
$1 \leq \tau \leq 2048$ minutes and $1 \leq r \leq 10$ (see figure (\ref{fig2}A)).
Outside this domain
in the ($r,\tau$) plane, the linear representation approximation breaks down.
Inside the domain of validity of the linear representation approximation, the 
spectrum of the generator is presented in figure (\ref{fig1}B) (resp. (\ref{fig2}B)) in the case of
FX DM/$\$$ (resp. CAC40).
The function $r \mapsto \zeta_r$ is in both cases 
(non trivially) concave, which by definition (see \cite{FP}) shows that the system has a
multifractal behavior.

\begin{figure}[tbhp]
\centerline{\epsfxsize=12truecm\epsfbox{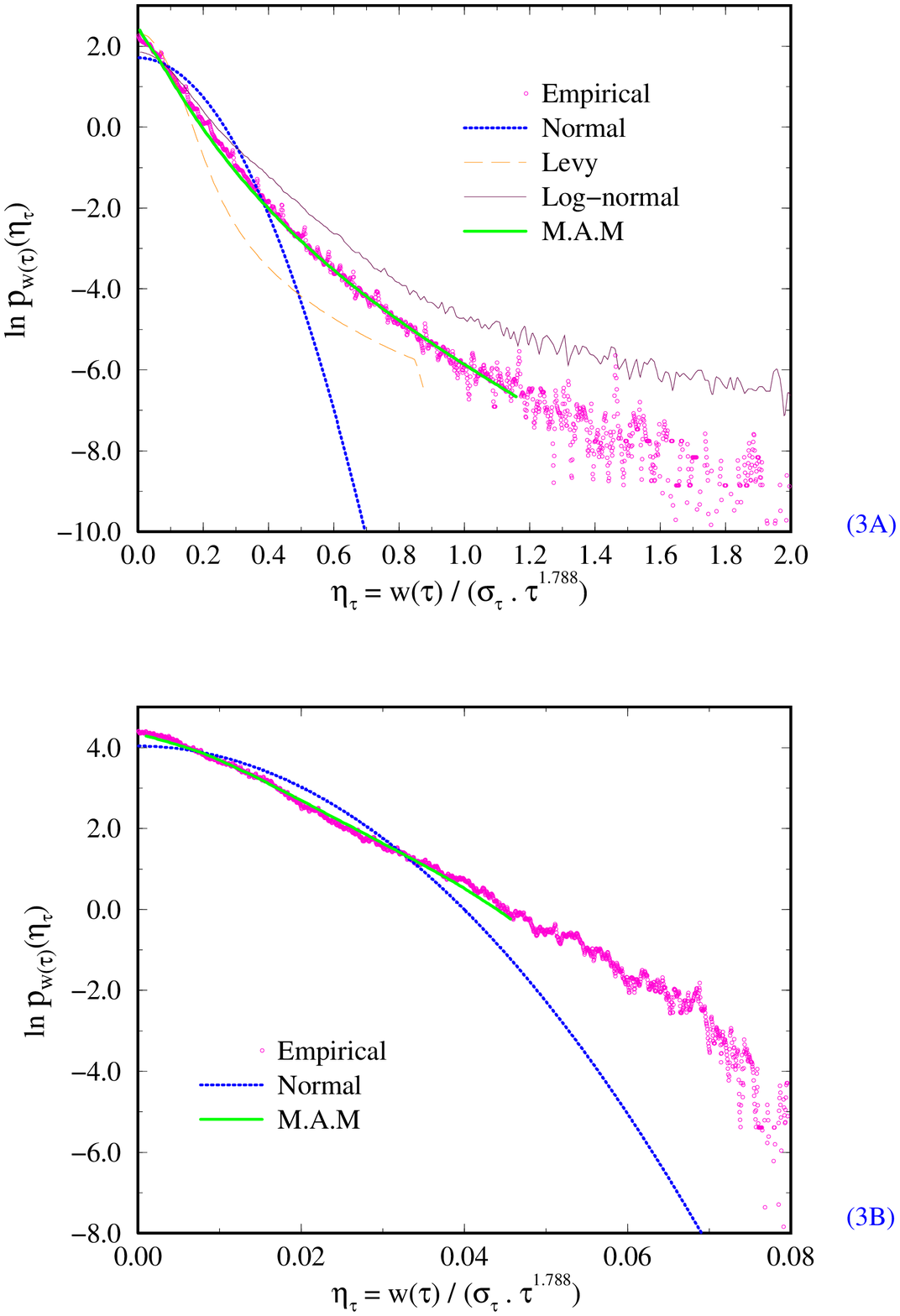}}
\caption{(3A): Presentation of the probability density function at
$\tau = 8$ minutes for FX DM/$\$$ index and comparison with empirical data.
(3B): Same presentation at $\tau = 512$ minutes for FX DM/$\$$ index.}
\label{fig3}
\end{figure}

\begin{figure}[tbhp]
\centerline{\epsfxsize=12truecm\epsfbox{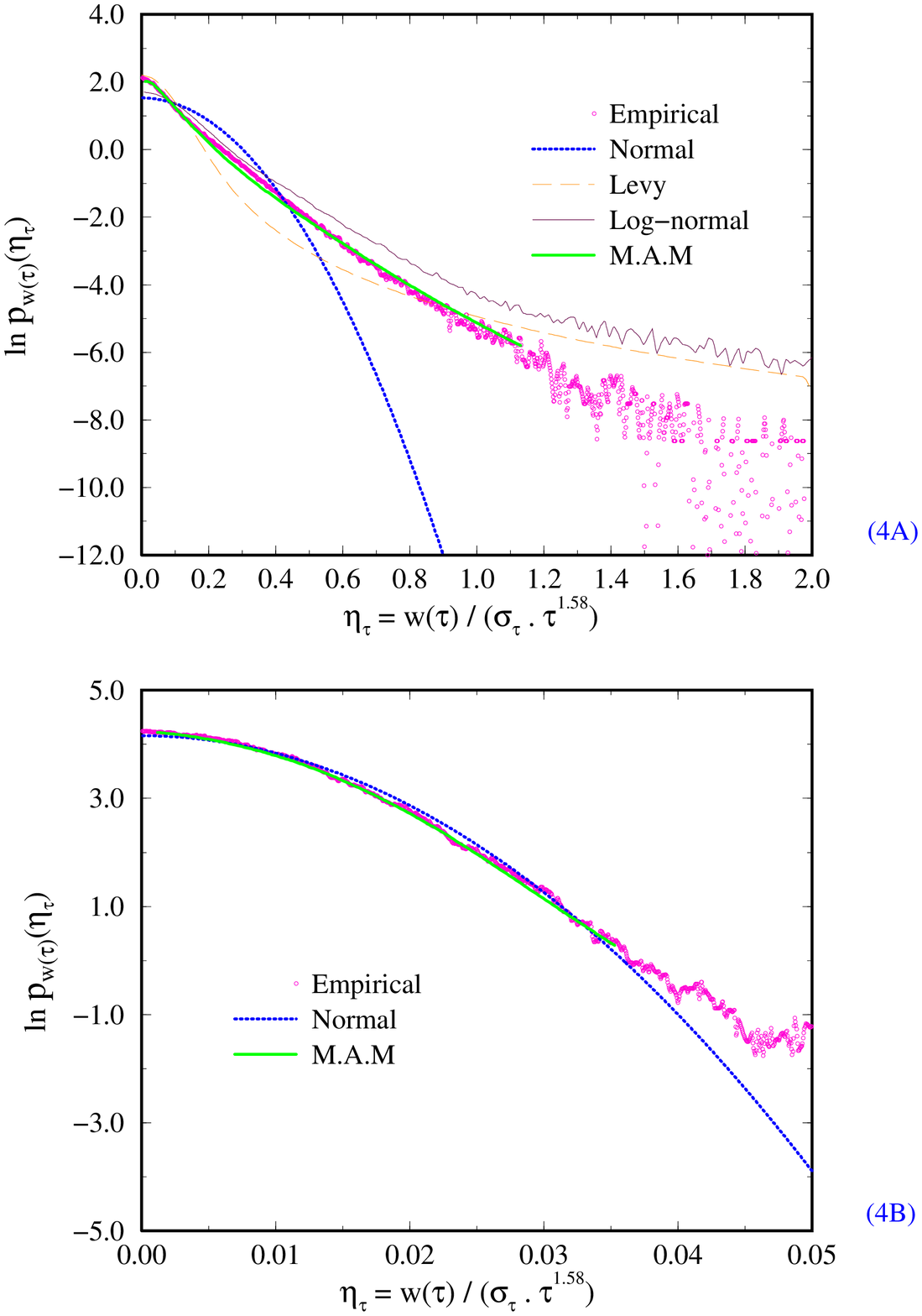}}
\caption{(4A): Presentation of the probability density function at $\tau = 8$
minutes for CAC40 index and comparison with empirical data.
(4B): Same presentation at $\tau = 512$ minutes for CAC40 index.}
\label{fig4}
\end{figure}

Finally, we present in figure (\ref{fig3}A) and figure (\ref{fig3}B) (resp. figure (\ref{fig4}A) and 
figure (\ref{fig4}B)) probability densities (M.A.M) given by (\ref{uap7}) and (\ref{uap8}), for $\tau=8$
 minutes and $\tau = 512$ minutes in the case of FX DM/$\$$ index (resp. CAC40). In all the cases, the experimental probability distribution is well 
approximated, for a large range of price increments, by the corresponding 
probability distributions in the family $\{ p_{w(\tau)} \}_{\tau>0} $ 
constructed by the inverse method developed in \cite{Br.Tch} and \cite{theseJMT}. 
Other commonly used probability
 distributions are also presented in the figures for illustration.

\end{document}